# A Comprehensive Content Verification System for ensuring Digital Integrity in the Age of Deep Fakes

Ravi Kanth Kaja
Microsoft
ravi.kaja@microsoft.com

*Abstract* – In an era marked by the widespread sharing of digital content, the need for a robust content-integrity verification goes beyond the confines of individual social media platforms. While verified profiles (such as blue ticks on platforms like Instagram and *X*) have become synonymous with credibility, the content they share often traverses a complex network of interconnected platforms, by means of re-sharing, re-posting, etc., leaving a void in the authentication process of the *content* itself. With the advent of easily accessible AI tools (like DALL-E, Sora, and the tools that are explicitly built for generating deepfakes & face swaps), the risk of misinformation through social media platforms is growing exponentially. This white paper discusses a solution, a *Content Verification System*, designed to authenticate images and videos shared as posts or stories across the digital landscape. Going beyond the limitations of blue ticks, this system empowers individuals and influencers to validate the authenticity of their digital footprint, safeguarding their reputation in an interconnected world.

This solution serves as a vital defense against the increasing risk posed by deepfakes generated by advanced AI. Just as *X* seeks to combat the issue of bots on its platform, this Content Verification System is intended to counter the proliferation of misinformation through morphed, forged, and deepfake content that circulates across the internet.

## 1 PROBLEM STATEMENT

In 2022, the digital realm witnessed a surge in deepfake incidents, from a deceptive video of Ukrainian President Volodymyr Zelensky urging soldiers to lay down arms to a misleading clip of US President Joe Biden calling for a national draft. Even Russian President Vladimir Putin and Hollywood star Tom Cruise fell victim to manipulated videos, adding to the growing concern.

Fast forward to 2024, and the dark side of AI manifested in sexually explicit deepfake images of Taylor Swift circulating on social media, sparking heated discussions among privacy advocates. On the other side, in a first-of-its-kind incident, a multinational company's Hong Kong office suffered a staggering financial loss of HK$200 million due to a sophisticated deepfake scam.

As artificial intelligence continues to advance rapidly, the realism of deepfakes is poised to increase, making it progressively challenging to distinguish them from authentic content. Deepfakes have the potential to manipulate public sentiments, tank stock prices, influence voters, or even create public unrest. A system to validate the authenticity and integrity of the online content is the need of the hour.

## 2 MAIN CONTENT

### 2.1 PROPOSED SOLUTION

Ongoing research is actively exploring the identification of deepfakes through Machine Learning, Generative Adversarial Networks (GAN), behavioral sciences, and by considering the existing constraints of generative AI. However, the continual advancements in AI are expected to heighten the difficulty of achieving success with these approaches. Further, a detector model would eventually help train better generators.

The alternative solution proposed here is to develop a *content verification system* in which

- An individual or an entity submits their content on the content verification platform
- The recipients of the content can verify the authenticity and the integrity of the content

Content in this context refers to images, audio or videos intended to be shared on social media platforms or the internet.

The proposed solution operates independently of individual social media platforms for validating the *integrity* of a *content being circulated* against the *original version*. For browser-based experience, an interface, either as a *native feature of the browser*, or through a *browser extension* can be implemented that serves 2 purposes.

- First, it enables users to submit the content to a trusted Register/Database/Ledger while submitting/posting it on the social media platform. (From here on referred to as Content Registration)
- Second, it facilitates the display of verification tags on content that has been verified (as shown below).

While a dedicated web application (or) a mobile app can serve as an interface to submit/verify the content, a more integrated feature that is closely embedded with the mobile platforms or native social media apps will be an attractive solution for users to submit/verify the content while posting/viewing it.

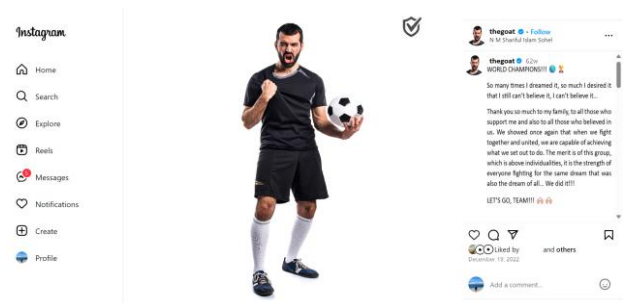

[Image by luis_molinero on Freepik]

Figure 1: An example social media post with content verification tag.

*Refer to the*  *verification tag on the top-right corner of the image*

### 2.1.1 TYPES OF CONTENT IN SCOPE

*Images, Audio & Video* are the 3 types of content which could benefit from a content verification system. In this paper, images have been taken under scope, but the solution proposed can be implemented for audio and video content as well.

In the case of video content, scene change detection can be achieved by comparison of localized histograms in between scenes, and scene averages can be calculated. The same solution that is implemented for images can be extended by applying it on scene averages.

*Note: For the rest of this paper, Content & Image will be used interchangeably*

### 2.1.2 LIMITATIONS OF DIGITAL SIGNATURE AND PROPOSED ALTERNATIVE

The default solution of *digital signatures* for the content is not applicable in this scenario as:

1. **PROBLEM:** The metadata (like .exif) of the image is stripped as soon as it is uploaded onto social media making it extremely difficult to maintain a *persistent identity* for the image while it is shared or exchanged across users and platforms.

   **SOLUTION:** An identity is generated for the image when it is onboarded to content verification system and this information is embedded into the image using *steganographic techniques*. The image's identity can be retrieved at any later point by extracting the embedded information.

   *Image steganography* is a technique of hiding data or information within an image in such a way that it is concealed from plain view.

2. **PROBLEM:** When an image is shared across social media, it goes through multiple transformations like compression, change of brightness, format change, resize, cropping, etc. Hence, validating the *integrity* of the image by strict comparison of the contents of the image (or) by their digital signature would lead to false positives (or true negatives) and is not an acceptable solution.

   **SOLUTION:** Image authentication & integrity check implementation using s*elective authentication*.

   *Selective Authentication* is a method used to verify the authenticity of images while providing robustness against any specific and desired manipulations (like compression, different filtering algorithms, etc) but also detecting any malevolent operations.

### 2.1.3 DEFINING AUTHENTICITY AND TRUST FOR THE CONTENT

A content trending over internet, or major social media platforms (in this context), can be questioned for its authenticity by asking these 3 questions:

1. *What* is the content, has it been tampered with, morphed or manipulated?

   When a user submits their content onto the platform, this verification system will be able to address that question with *confidence score* (more details follow in the next sections).

2. *Who* is this content from? While the content might land on one's social media account after multiple hops, the question always is - is this some miscreants playing with the public sentiments or is it from a genuine source.

   If there is a message intended to be shared with their public by the US President Joe Biden, or Ukrainian President Volodymyr Zelensky (like mentioned in the Problem Statement), it will originate from their verified profiles.

   When a user submits their content, their *social media identity* can be *tagged against the content* in the verification platform. Further verification of the user (profile) can be achieved through the '*verified profiles*' feature of the respective platform.

The necessity of this feature arises from the fact that there is currently no feature to tie the content to its source (even as an optional feature) & content often originates from one platform but eventually spreads across multiple platforms.

3. *Where* does this content originate from. Is the source, a hardware device like a camera, or a generative AI model.

There is active work in progress, in the field of content provenance, both from generative AI models and embedded systems (like cameras). The ability to embed this information, when some content is submitted to the platform, gives the users *additional context on the source of information*, along with its *integrity*.

In this paper, we primarily focus on the 'What' part – validating the integrity of a content seen on a user's feed/inbox against the initial content that has been saved on the verification platform & published (for the first time) by the 'assumed' content owner. As this is achieved through a content repository, the 'Who', 'Where' & 'When' details can be attributed as metadata against the content for additional context.

## 2.2 CONTENT VERIFICATION SCHEME

### 2.2.1 CONTENT REGISTRATION

1. Extract Features of the Image. *Features* refer to the characteristics or attributes that describe the image content. A frequency-based signature of the image derived from its Discrete Cosine Transform is one such example.
2. Save it in the database/ledger of Content Verification System
3. Create a Barcode Encoded with the address of the image in the database
4. Embed the Barcode as an invisible watermark into the Image
5. Return the Image for sharing across social media/internet

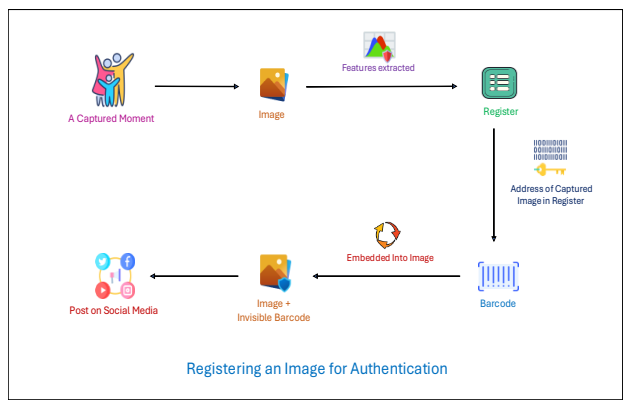

Figure 2: Content Registration Scheme explained

### 2.2.2 CONTENT VERIFICATION

1. Extract the barcode/watermark from the Image
2. Decode the barcode to get the address in the database for the Image
3. Fetch the original image's features from the database/Register
4. Extract the features from the image under verification
5. Compare the features and return a confidence score on the integrity of the image under verification. Selective Authentication techniques are to be employed here to stay tolerant of transformations like image compression, that are bound to happen on social media platforms.

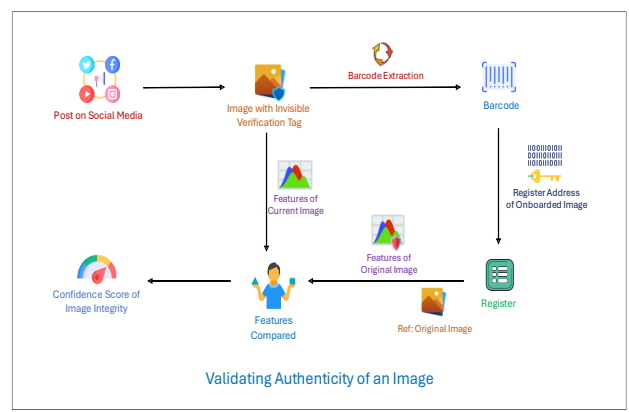

Figure 3: Content Verification Scheme explained

## 2.3 DESIGN CONSIDERATIONS

### 2.3.1 CONTENT REGISTRATION

#### 2.3.1.1 FEATURE EXTRACTION

An approach leveraging Discrete Cosine Transform is employed for feature extraction and formulation of image signature. Discrete Cosine Transform (DCT) is a mathematical technique used in image processing and signal compression, in which an image is represented by a matrix of coefficients representing the amplitude distribution of cosine waves which constitute the final image. Various techniques have been evaluated based on the report provided by Adil Haouzia & Rita Noumeir in 'Methods for image authentication: a survey ', and DCT has been chosen, as it stands out amongst the rest, in its robustness against image modifications like jpeg compressions, filtering, brightening etc. The algorithm employed involves breaking down an image into 8x8 pixel blocks and performing DCT on each individual block to have a more localized feature definition for each individual block.

```python
def perform_dct(img, blk_size):

    image_size = img.shape
    dct = np.zeros(image_size)

    #Perform blockSize x blockSize DCT on image
    for i in r_[:image_size[0]:blk_size]:
        for j in r_[:image_size[1]:blk_size]:

            #Adjust mid-grey to zero in 8-bit image
            blk_adj = np.ones((blk_size,blk_size))*128
            mid_adj = (img[i:(i+blk_size),j:(j+blk_size)]-blk_adj)
            #Normalize the block [0,1]
            normalized_blk = np.float32(mid_adj)
            #DCT using OpenCV2 & Assigning to numpy dynamic array element
            dct[i:(i+blk_size),j:(j+blk_size)] = cv2.dct(normalized_blk)

    return dct
```

Figure 4: Python code for Feature Extraction through Discrete Cosine Transform (DCT)

### 2.3.1.2 IMAGE SCALING ACROSS SOCIAL PLATFORMS

As image manipulations are common for any content when it is exchanged across social media platforms, a feature definition of one dimension might not always match with the feature definition of a scaled-up/scaled-down image. Below is a reference to some of the standard dimensions & aspect ratios of different social platforms as of today. When an image is onboarded for verification, its dimensions are recalibrated to the closest match of recommended aspect rations and the features are extracted from the resized images.

| Platform | Instagram | Facebook | X | LinkedIn |
|---|---|---|---|---|
| Profile Photo | 320 x 320 | 170 x 170 | 400 x 400 | 400 x 400 |
| Landscape | 1080 x 566 | 1200 x 630 | 1600 x 900 | 1200 x 627 |
| Portrait | 1080 x 1350 | 630 x 1200 | 1080 x 1350 | 627 x 1200 |
| Square | 1080 x 1080 | 1200 x 1200 | 1080 x 1080 | 1080 x 1080 |
| Stories/ Reels | 1080 x 1920 | 1080 x 1920 | - | - |
| Cover Photo | - | 851 x 315 | 1500 x 1500 | 1128 x 191 |

Table 1: Commonly used/supported dimensions for various content types across different social media platforms (As of January 2024)

### 2.3.1.3 CONTENT ID CREATION

A content ID which is used as a pointer in the database is created. As the content ID needs to be unique, a hashing algorithm is used for its generation. However, the size of content ID needs to be as minimal as possible, as its information needs to be embedded in the host image, and larger size of content ID would mean increased size of the host image, and the possibility of visible distortions in the host image due to increased energy of the embedded watermark.

### 2.3.1.4 STEGANOGRAPHY CONSIDERATIONS

The content ID generated in the above steps needs to be embedded into the host image through steganography to ensure that the visual experience for the host image remains intact post the addition of it. The following aspects and design choices have been taken into consideration to make this a robust solution:

- **FAST FOURIER TRANSFORM**: As the content is expected to go through multiple modifications across social platforms, frequency domain is chosen as a preference over spatial domain for the steganography process as it is more robust to the changes that are generally expected.
  Fast Fourier Transform (FFT) has been chosen for this process considering its resistance to certain steganalysis techniques. FFT transforms an image from its spatial domain (pixel values) into the frequency domain. In the frequency domain, an image is represented as a combination of different sinusoidal waveforms, each with its own frequency, phase, and magnitude.

```python
def embed_watermark(host, watermark, xmap, margins, alpha):

    if len(watermark.shape) == 2:
        watermark = np.repeat(watermark[:, :, np.newaxis], 3, axis=2)

    n_host = normalize_rgb(host)
    n_watermark = normalize_rgb(watermark)

    fft_host = np.fft.fft2(n_host, None, (0, 1))

    zero_filler_x = n_host.shape[0]//2-margins[0]*2
    zero_filler_y = n_host.shape[1]-margins[1]*2
    buffer = np.zeros((zero_filler_x, zero_filler_y, 3))
    buffer[:n_watermark.shape[0], :n_watermark.shape[1]] = n_watermark

    xh, xw = xmap[:2]

    fft_host[+margins[0]+xh, +margins[1]+xw] += buffer * alpha
    fft_host[-margins[0]-xh, -margins[1]-xw] += buffer * alpha

    ifft_host = np.fft.ifft2(fft_host, None, (0, 1))

    ifft_host = ifft_host.real
    ifft_host = np.clip(ifft_host, 0, 1)

    return ifft_host
```

Figure 5: Python code for watermarking the Content ID through Fast Fourier Transform (FFT)

- **ERROR CORRECTION CODING THROUGH 2D QR CODES**: To be able to successfully retrieve the embedded information from the watermarked image, an efficient Error Correction Coding is required. 2D QR Codes which have out of the box error correction coding capabilities have been chosen to address this requirement. The content ID is first encoded into a 2D QR Code with required error correction input, and the generated QR Code is watermarked into the host image. It is important to note that the higher the error correction code value, the more the energy of the QR Code, and the more the introduced distortions to the host image.

```python
def generate_qrcode(text):
    text = f'<{text}>'
    rs = RSCodec(10)
    encoded_text = rs.encode(text.encode('utf-8'))

    qr = qrcode.QRCode(
        version=1,
        error_correction=qrcode.constants.ERROR_CORRECT_Q,
        box_size=10,
        border=4,
    )

    qr.add_data(encoded_text)
    qr.make(fit=True)
    qr_img = qr.make_image(fill_color="black", back_color="white")

    return qr_img
```

Figure 6: Python code for QR Code generation with Error Correction

- **MASKING THE WATERMARK WITH CRYPTOGRAPHY:** A watermarked image should further be protected from attackers who attempt to tamper with the watermark and make it unfit for verification. To achieve this, the index and sequence of frequency coefficients of the host image which are used to embed the watermark needs to be unpredictable and unreadable. This is achieved by using a *Master Key* which is only accessible to the Content Verification Platform and generating random numbers using this Master Key as a seed and creating a pattern of indexes from this random numbers, which are used to embed the watermark into the host image.

```python
def generate_xmap(shape, key = None):
    xh = np.arange(size[0])
    xw = np.arange(size[1])

    if key:
        random.seed(key)
        for i in range(shape[0], 0, -1):
            j = random.randint(0, i)
            xh[i-1], xh[j] = xh[j], xh[i-1]

        for i in range(shape[1], 0, -1):
            j = random.randint(0, i)
            xw[i-1], xw[j] = xw[j], xw[i-1]

    xh = xh.reshape((-1, 1))

    return xh, xw
```

Figure 7: Python code for generating indices cryptographically for watermark embedding

### 2.3.2 CONTENT VERIFICATION

#### 2.3.2.1 QR Code Cleanup

A watermark extracted from the host image often comes with some distortions and noise, especially when the host image has been altered in its dimensions or brightness. While the error correction coding helps in maintaining the structural integrity of the QR Code, *denoising* the image, and applying *thresholding* on the extracted watermark has been implemented to get a cleaned image which is more readable for any QR Code decoding.

#### 2.3.2.2 Feature Extraction in Grey Scale

For the sake of simplicity, feature extraction for image verification is performed on the grey scale versions of the original image and the image requiring verification. However, the same approach can be extrapolated by performing feature extraction and comparison across each of the 3 channels, for luminous intensity of Red, Blue and Green if required.

#### 2.3.2.3 Feature Comparison Metrics

The following 2 metrics have been implemented for the feature comparison, and to assess the similarity between the original image, and image under verification.

- **MEAN SQUARED ERROR (MSE):** MSE measures the average squared difference between pixel values in two images. A lower MSE indicates that the two images are more similar or have less distortion, while a higher MSE suggests greater dissimilarity or more significant distortion.
- **PEAK SIGNAL-TO-NOISE RATIO (PSNR):** PSNR measures the ratio of the peak signal level (the maximum possible pixel value) to the MSE between two images. PSNR is measured in decibels, and a higher PSNR value indicates better image quality and less distortion.

## 2.4 EXPERIMENT RESULTS

The *Framework* for *Content Verification System* has been tested with different scenarios, the results, and some of the key findings have been documented below.

### 2.4.1 SCENARIO: CONTENT REGISTRATION

A host image with the tag 'Original Content' in the below picture is taken as input, and a content ID generated for it is encoded into a 2D QR Code with error correction applied, and the resultant QR Code is embedded into the host image, the output of which is the image with the tag 'Watermarked Content' shown below.

*Results:* Watermarked Image is generated with no observable distortions, maintaining similarity to its original image.

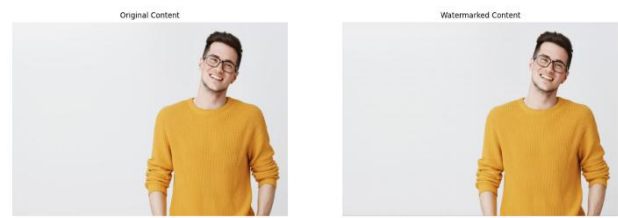

[Image by cookie_studio on Freepik]

Figure 8: Images: Original & Watermarked

### 2.4.2 SCENARIO: CONTENT VERIFICATION OF A GENUINE IMAGE

An image with the tag 'Validation Content' in the below picture is taken as input, and is validated against its reference, the original content which is tagged as 'Registered Content' in the below picture. DCT is performed for 8x8 pixel blocks of grey scale images, and they are compared with MSE & PSNR metrics.

*Results:* DCT derived from both images look visually identical. MSE & PSNR results support this observation.

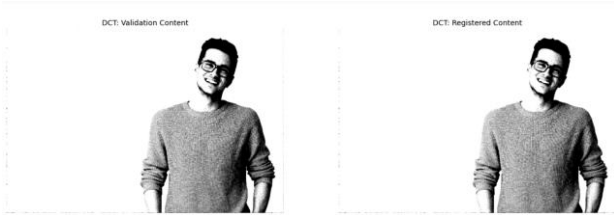

[Image by cookie_studio on Freepik]

Figure 9: Content Verification Images: Image for Validation & Registered Image

```
Mean Squared Error: 0.0
Peak Signal-to-Noise Ratio: inf
```

### 2.4.3 SCENARIO: CONTENT VERIFICATION OF SLIGHTLY/MODERATELY MORPHED IMAGE

An image with the tag 'Validation Content' in the picture below is taken as input, and is validated against its reference, 'Registered Content'. The 'Validation Content' is *morphed*, by *face-swapping a new face* into the original image.

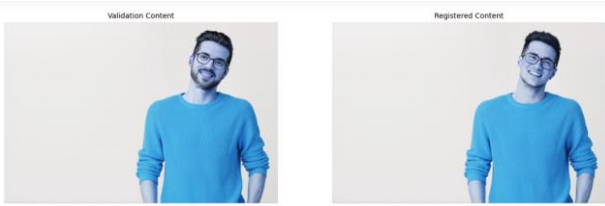

[Image by cookie_studio on Freepik]

Figure 10: Face-Swapped Image & Original Image

DCT is performed for 8x8 pixel blocks of grey scale images, and they are compared with MSE & PSNR metrics.

*Results:* DCT derived from both the images have visible distinctions for the facial features. Higher MSE & relatively lower PSNR corroborate with this dissimilarity.

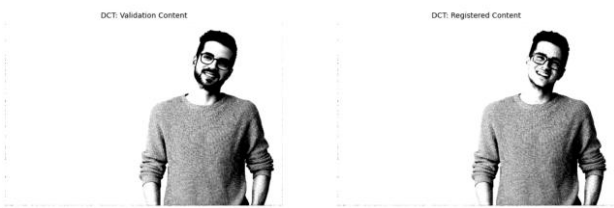

[Image by cookie_studio on Freepik]

Figure 11: Content Verification Images: Face-Swapped Image & Original Image

```
Mean Squared Error: 115.5239
Peak Signal-to-Noise Ratio: 27.5040
```

### 2.4.4 SCENARIO: CONTENT VERIFICATION OF HIGHLY MORPHED IMAGE

An image with the tag 'Validation Content' in the picture below is taken as input, and is validated against its reference, 'Registered Content'. The 'Validation Content' is morphed, by *adding a new person* into the original image.

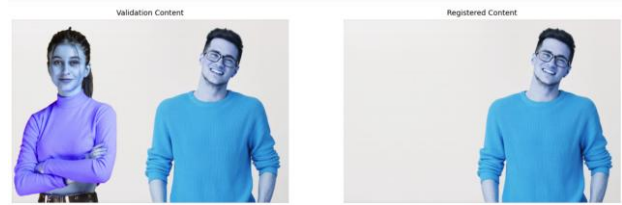

[Image by cookie_studio on Freepik]

Figure 12: Morphed Image & Original Image

DCT is performed for 8x8 pixel blocks of grey scale images, and they are compared with MSE & PSNR metrics. Significantly high MSE & significantly low PSNR imply the high levels of distinction between the 'Validation Content' and the 'Registered Content'.

*Results:* DCT derived from both the images have clear visible distinctions

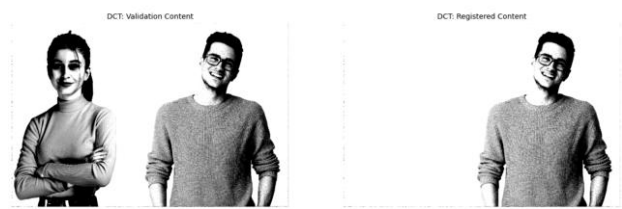

[Image by cookie_studio on Freepik]

Figure 13: Content Verification Images: Morphed Image & Original Image

```
Mean Squared Error: 2151.5111
Peak Signal-to-Noise Ratio: 14.8033
```

### 2.4.5 SCENARIO: IMPACT ON WATERMARK WHEN IMAGE IS BRIGHTENED

The image is brightened, introducing high energy changes to the frequencies of the image, including the frequency coefficients where watermark is embedded. Post-extraction operations of cleaning up the extracted watermark by performing denoising and thresholding are performed.

*Results:* The denoised QR Code was successfully decoded

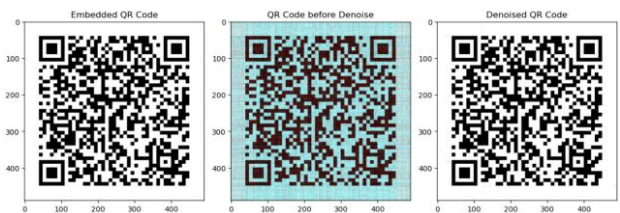

Figure 14: QR Codes: Embedded, extracted (Before Denoising), extracted (After Denoising)

## 2.5 FUTURE WORK & IMPROVEMENTS

The results from the experiment were promising, and the results recommend the following areas of improvement and exploration:

1. Improving efficiency in watermarking with lesser impact to the size of the original host image.
2. Certain transformations to the original image have been found to distort the watermark making it difficult to extract it. A redundant system, where watermarks are embedded at different frequencies, or where they are embedded in both spatial and frequency domains can make the image identification through watermarks more robust.
3. Mean Squared Error and Peak Signal-to-Noise Ratio have been observed to be more sensitive to certain changes to the images like brightness and filters. A redundant feature comparison system, which covers both spatial and frequency domains, might be more useful and needs exploration.
4. There is scope for exploring alternatives to QR Code based watermarking to achieve error correction coding, which might be less in energy and can have lesser (or negligible) distortions to the original host image.
5. Evaluation of technologies/solutions for the storage of the submitted content that addresses aspects like scale, performance and user-privacy.

## 2.6 INDUSTRY LANDSCAPE, RESEARCH AND STANDARDS – CAI & C2PA

The *Content Authenticity Initiative (CAI)* was founded by Adobe in 2019, with partners like the New York Times & Twitter with a goal to create open-source tools & foster a community to support digital content authenticity. The *Coalition for Content Provenance and Authenticity (C2PA)* was formed in 2021 by *Adobe*, *Microsoft, BBC*, and other major tech companies to create technical standards for content provenance and authenticity. Other industry leaders such as *Google* and *Meta* have since joined the effort.

C2PA provides the end-to-end open technical standards that CAI tools and other systems use to ensure content authenticity. By attaching verifiable information about the origin and history of digital content, they make it easier to trace the source and modifications of digital media, thereby providing a way to validate the authenticity of the information.

### 2.6.1 CONCEPTUAL DISTINCTIONS FROM C2PA

#### FOCUS ON METADATA VS WATERMARKING

C2PA primarily focuses on developing standards and specifications for attributions and metadata association with the content, wherein the provenance information sits along with the content. Further, watermarking is chosen as a redundancy option, in the event, the metadata is stripped from the content, the watermark can be used to point to a persistent datastore pointing to the Provenance records of the said content.

*In contrast*, this paper's approach solely relies on watermarking an identifier to the content that would point to the information stored in a repository.

#### VERIFIABLE ATTRIBUTES VS CONFIDENCE SCORE

Other C2PA tools offer verifiable attributes of source and modification details over content.

*In contrast*, this paper's fundamental proposal focuses on providing a confidence score by comparing the original content with the content on the feed through feature comparison. This method can simplify the data privacy compliance aspects (as the source & modification information is not in picture, except for social media profile linking), but at the cost of reduced information and context.

#### OPEN-SOURCE WATERMARKING VS CRYPTOGRAPHIC APPROACH

In one of the papers published by the CAI group - 'To Authenticity, and Beyond! Building Safe and Fair Generative AI Upon the Three Pillars of Provenance Watermarking' an opensource watermarking algorithm is chosen, meaning anyone with this knowledge can strip or add a watermark. While perceptual hashing of the image is intended to prevent spoofing in this approach, some of the perceptual hashing algorithms have been proven to be vulnerable to hash collisions and hacks. Although, it is worth noting that the algorithm used in the above-mentioned paper, hasn't been particularly cited to have such vulnerability so far.

*In contrast*, this paper suggests leveraging a cryptographic approach for embedding watermarks, making it impractical for malicious actors to remove or modify the watermark. Further exploration is due on topics related to secret management, secret exchange and usage of privacy-preserving cryptographic techniques.

### 2.6.2 CONCEPTS PLANNED FOR ADAPTATION FROM C2PA & CAI IN FUTURE WORK

1. C2PA has established comprehensive standards for Content Credentials and Attributions. We plan to integrate C2PA's schema and specifications while further investigating a robust data privacy framework.
2. Watermarking techniques utilized in CAI tools employ AI models to ensure minimal disruption to the original content. Combining these methods with additional obscurity measures for enhanced security presents a promising avenue for future development.
3. The Content Credentials from C2PA establish the integrity of the attributes assigned to a content through digitally signed claims. However, to validate the authenticity of the 'Identity Attribute' (for example, a username from a renowned Identity Provider like Azure AD or Google Identity Platform) itself, there is a construct called Verifiable Credentials, which allows for attested credentials from Identity Providers to be included in the Provenance Information. This paper aims to leverage the same approach for tying up the content with its creator.

## 3 CONCLUSIONS

In conclusion, this paper underscores the critical need for a robust content-integrity verification system in the era of widespread digital content sharing. This system empowers individuals and influencers to affirm the authenticity of their digital footprint, fortifying their reputation in an increasingly interconnected world.

As a significant defense against the rising menace of deep fakes, this framework not only provides a practical solution but also serves as a proactive measure against the potential proliferation of morphed, forged, and manipulated content on the internet. Much like the proactive stance taken by platforms like X in combating bots, a Content Verification System can contribute to the ongoing battle against the spread of deceptive content, safeguarding the integrity of shared media.

While the inclusion of code snippets for the implemented framework offer a tangible demonstration of the system's functionality, the references to complimenting work that is happening in the industry (with C2PA & CAI), and other alternative technologies provide valuable insights into the broader landscape of content-integrity solutions. This paper also highlights the vast scope for further development and refinement of the proposed Content Verification System.

As we navigate the evolving challenges of the digital age, the pursuit of innovative solutions remains paramount, and this work represents a sincere effort in the ongoing efforts to maintain trust and authenticity in the digital landscape.